\documentclass[12pt,preprint]{aastex}

\shorttitle{NGC 288 Abundances}
\shortauthors{Hsyu, Johnson, \& Rich}

\begin{document}

\title{Light Element Chemistry and the Double Red Giant Branch in the Galactic 
Globular Cluster NGC 288}

\author{
Tiffany Hsyu\altaffilmark{1},
Christian I. Johnson\altaffilmark{2,4}, 
Young--Wook Lee\altaffilmark{3}, and
R. Michael Rich\altaffilmark{1}
}

\altaffiltext{1}{Department of Physics and Astronomy, UCLA, 430 Portola Plaza,
Box 951547, Los Angeles, CA 90095--1547, USA; tiffanyhsyu@ucla.edu; 
rmr@astro.ucla.edu}

\altaffiltext{2}{Harvard--Smithsonian Center for Astrophysics, 60 Garden
Street, MS--15, Cambridge, MA 02138, USA; cjohnson@cfa.harvard.edu}

\altaffiltext{3}{Center for Galaxy Evolution Research and Department of 
Astronomy, Yonsei University, Seoul 120--749, Korea; ywlee2@yonsei.ac.kr}

\altaffiltext{4}{Clay Fellow}

\begin{abstract}

The globular cluster NGC 288 was previously reported to exhibit two distinct
red giant branches (RGBs) in the narrow--band Calcium (HK) and Str{\"o}mgren b
and y band passes.  In order to investigate this phenomenon further, we 
obtained moderate resolution (R$\sim$18,000) spectra of 27 RGB stars in NGC 288
with the Hydra multifiber spectrograph on the Blanco 4m telescope at Cerro
Tololo Inter--American Observatory.  From these data we derive iron 
($\langle$[Fe/H]$\rangle$=--1.19; $\sigma$=0.12), oxygen 
($\langle$[O/Fe]$\rangle$=$+$0.25; $\sigma$=0.13), and sodium
($\langle$[Na/Fe]$\rangle$=$+$0.15; $\sigma$=0.26) abundances using standard
equivalent width and spectrum synthesis techniques.  Combining these data
with those available in the literature indicates that the two giant branches
have distinctly different light element chemistry but do not exhibit a 
significant spread in [Fe/H].  A new transmission tracing for the CTIO Ca
filter, obtained for this project, shows that CN contamination is the primary 
spectral feature driving the split RGB.  Interestingly, the CN leak in the 
current CTIO Ca filter may be used as an efficient means to search for
CN--weak and CN--strong stars in systems with otherwise small Ca abundance
variations.

\end{abstract}


\section{INTRODUCTION}

It has long been known that significant variations in light element abundances 
exist within almost all globular clusters, and in fact may be a defining 
characteristic that separates globular clusters from other stellar populations 
(e.g., see reviews by Gratton et al. 2004; 2012 and references therein).
Recently, these light element variations have been linked to multiple 
photometric sequences identified by the careful selection of color indices that
exploit differences in spectral features between stars, based on their 
composition (e.g., Monelli et al. 2013; Piotto et al. 2013; see also reviews 
by Piotto et al. 2009; Gratton et al. 2012).  In particular, this has led to 
the discovery that most globular clusters experienced at least two distinct 
star formation events.  In this scenario, a first generation of stars forms 
from gas primarily polluted by core--collapse supernovae (SNe) and with a 
composition nearly identical to metal--poor halo field stars.  Subsequently, 
before all of the gas is lost from the proto--cluster, a second generation 
forms near the cluster center with the same [Fe/H] abundance but exhibiting 
significantly different light element (e.g., O--poor and Na--rich) chemistry 
(e.g., D'Ercole et al. 2008; Valcarce \& Catelan 2011).  The pollution source 
and formation time scale for the second generation stars are not presently 
well--constrained, but the leading candidates include intermediate mass 
asymptotic giant branch (AGB) (e.g., Ventura \& D'Antona, 2008; Karakas 2010), 
rapidly rotating massive main--sequence stars (e.g., Decressin et al. 2007),
massive binary stars (e.g., de Mink et al. 2009; Izzard et al 2013), and 
possibly super--massive stars (e.g., Denissenkov \& Hartwick 2014).

While the small ($<$0.1 dex) [Fe/H] and [$\alpha$/Fe] dispersion and typically
(inferred) moderate He abundance difference between first and second generation
stars in most globular clusters seems to exclude SNe as a significant 
contributor to the gas from which second generation stars form, Roh et al.
(2011) found photometric evidence that NGC 288 stars may exhibit a 
non--negligible spread in metallicity.  The possible metallicity spread is 
evidenced by a split red giant branch (RGB), observed in the y versus 
hk\footnote{The hk index is defined in Roh et al. (2011) as 
hk=[(Ca--b)--(b--y)], where Ca, b, and y are the CTIO Ca and Str{\"o}mgren b
and y filters.} color--magnitude diagram, that may be best fit by isochrones
in which second generation stars are $\sim$1.5 Gyr younger, He--enhanced 
($\Delta$Y$\approx$0.03), and more metal--rich ($\Delta$[M/H]$\sim$0.16 dex)
than the first generation stars.  The observed split RGB in NGC 288 when using
the hk index follows similar observations by Lee et al. (2009) of other 
Galactic globular clusters, and suggests there may be a link between SNe and 
the well--established light--element variations mentioned previously.  However,
Carretta et al. (2010) investigated possible [Ca/H] variations via 
spectroscopy in a sample of 200 RGB stars in 17 globular clusters, including 
NGC 288, and did not find any significant correlations between [Ca/H] and 
light--element chemistry.  Consequently, there is a disconnect between the 
observed spread in RGB sequences produced by the hk index photometry and 
direct abundance measurements from high--resolution spectroscopy.

In this paper we use high--resolution spectroscopy of RGB stars in NGC 288 to 
test whether any [Fe/H] or light element chemistry differences exist between 
stars on the two giant branches.  We also provide a new transmission curve for 
the CTIO Ca narrow--band filter, discovering that the apparent population 
dispersion of NGC 288 was caused by the filter bandpass covering the CN band 
near 3885 \AA\ and is not due to a dispersion in metallicity.

\section{OBSERVATIONS, TARGET SELECTION, AND DATA REDUCTION}

The observations for all RGB stars in NGC 288 were taken on 2011 September
9--11 at the Cerro Tololo Inter--American Observatory (CTIO) using the Hydra
multifiber spectrograph on the Blanco 4m telescope.  Targets were selected
from the RAdial Velocity Experiment (RAVE) database (Siebert et al. 2011), and
only stars with radial velocities within $\pm$10 km s$^{\rm -1}$ of the 
--46.4 km s$^{\rm -1}$ cluster velocity (Dinescu et al. 1999) were considered
potential members.  Cluster membership was also confirmed using \emph{fxcor}
in IRAF\footnote{IRAF is distributed by the National Optical Astronomy
Observatory, which is operated by the Association of Universities for Research
in Astronomy, Inc., under cooperative agreement with the National Science
Foundation.}.  Our derived radial velocities for NGC 288 stars are in agreement
with the RAVE values.  We find an average heliocentric radial velocity of 
--43.5 km s$^{\rm -1}$ ($\sigma$=4.1 km s$^{\rm -1}$), with individual values 
listed in Table 1.

\begin{figure}
\epsscale{1.00}
\plotone{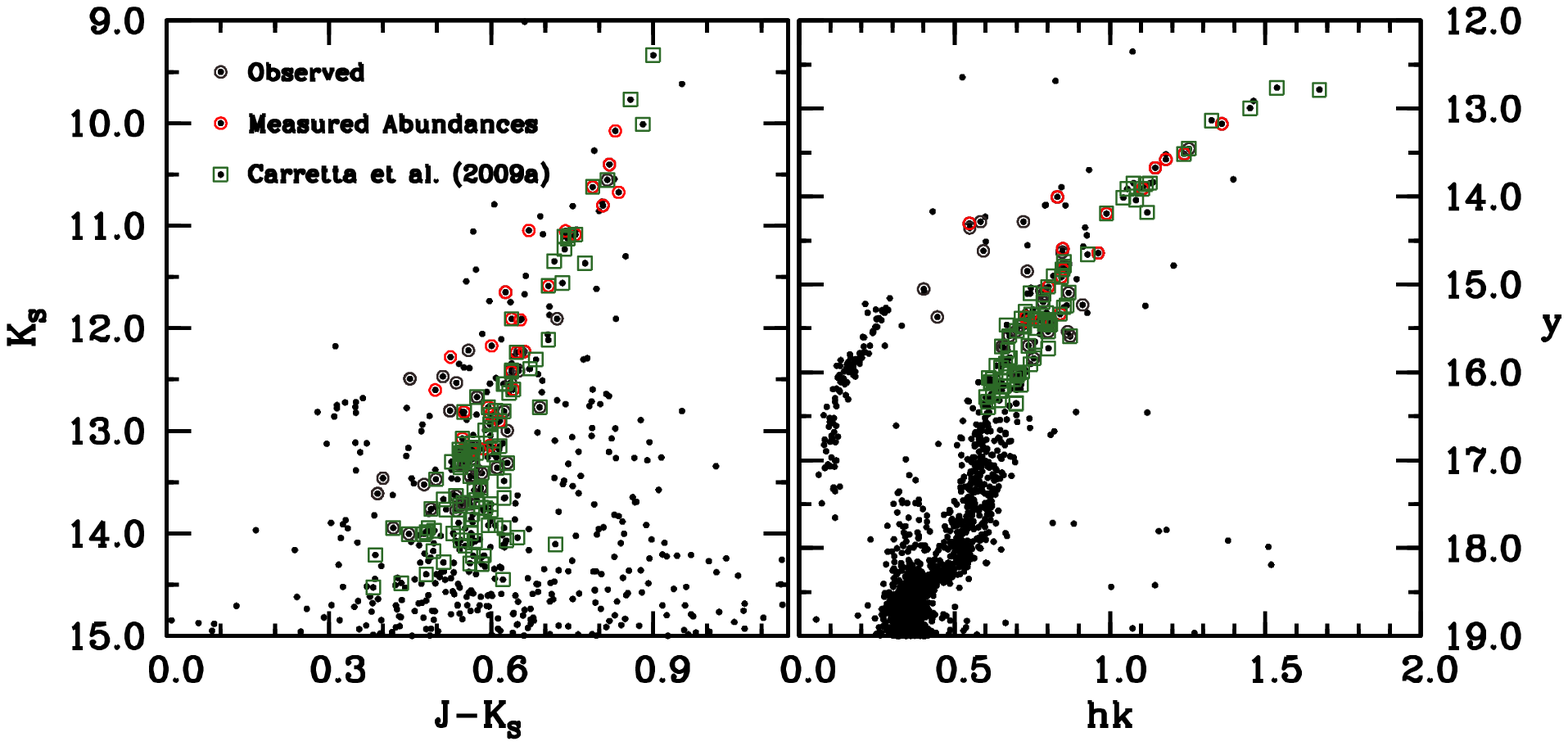}
\caption{\emph{left:} A 2MASS K$_{\rm S}$ versus J--K$_{\rm S}$
color--magnitude diagram illustrating the targets observed in the present work 
(open grey circles), those for which the S/N was high enough to derive 
abundance ratios (open red circles), and those observed in Carretta et al.
(2009a; open green boxes).  \emph{right:} A similar color--magnitude diagram
using the hk and y--band photometric data from Roh et al. (2011).  Note the 
double RGB.}
\label{f1}
\end{figure}

Coordinates and photometry for potential members were taken from Lane et al.
(2011) and 2MASS (Skrutskie et al. 2006).  We utilized two different Hydra
configurations to obtain spectra for $\sim$50 RGB stars with luminosities
ranging from the level of the horizontal branch to the RGB--tip.  
Color--magnitude diagrams illustrating the evolutionary state of all target
stars are provided in Figure \ref{f1}.  Unfortunately, we were not able to
fine--tune a configuration to obtain an equal number of stars on both the 
``blue" and ``red" RGBs shown in the right panel of Figure \ref{f1}.  A
single bench spectrograph setup was used for all observations.  The setup
included the use of the large (300 $\mu$m) fibers, 316 lines mm$^{\rm -1}$
Echelle grating, 400 mm Bench Schmidt camera, and E6257 filter to achieve a
resolving power of R($\lambda$/$\Delta$$\lambda$$\approx$18,000).  The 
spectra ranged from $\sim$6145--6350 \AA.

Basic data reduction was accomplished using the IRAF tasks \emph{ccdproc} and
\emph{dohydra} to trim the overscan region, bias subtract the images, 
identify and trace the fibers, correct for scattered light, remove cosmic
rays, apply the flat--field correction, linearize the wavelength scale based
on ThAr lamp exposures, and to subtract the sky spectra.  Telluric removal
was carried out using the \emph{telluric} IRAF task and observations of 
several rapidly--rotating B stars observed at various air masses.  The 
reduced images were then corrected for heliocentric velocity variations and
co--added with the \emph{scombine} task.  Unfortunately, the observing 
conditions during all NGC 288 observations were poor and included light clouds
and seeing of $\sim$1.5$\arcsec$.  This severely reduced the signal--to--noise 
ratios (S/N) of the spectra, leaving only 27 stars for which some 
abundances could be measured.  The final co--added spectra had typical S/N
of about 30.

\section{ANALYSIS}

\subsection{Model Stellar Atmospheres}

We utilized published photometry in the near--infrared J, H, and K$_{\rm S}$
bands from 2MASS to obtain effective temperatures (T$_{\rm eff}$) with the 
color--temperature relation described in Alonso et al. (1999; 2001).  In order
to use the Alonso et al. (1999; 2001) calibration, we converted the 
dereddened 2MASS J--K$_{\rm S}$ colors onto the Telescopio Carlos S\'{a}nchez
(TCS) system using the transformations summarized in Johnson et al. (2005).
We assumed E(B--V)=0.03 (Harris 1996; 2010 edition) for the color excess and
E(J--K$_{\rm S}$)/E(B--V)=0.527 Rieke \& Lebofsky 1985).  Surface gravity was
determined using the standard relation,
\begin{equation}
log(g_{*})=0.40(M_{bol.}-M_{bol.\sun})+log(g_{\sun})+4(log(T/T_{\sun}))+
log(M/M_{\sun}),
\end{equation}
with the bolometric magnitude calculated from Buzzoni et al. (2010) and 
assuming a mass of 0.8 M$_{\rm \odot}$.  Temperatures for the program stars
ranged from 4185 to 5215 K and the surface gravity ranged from 0.95 to 2.30 
cgs.

Individual model stellar atmospheres were created by interpolating within the
$\alpha$--enhanced (AODFNEW) ATLAS9 grid (Castelli et al. 1997).  Temperatures
and gravities were held fixed at their photometric values, and initial
estimates for metallicity and microturbulence (vt) were set at [Fe/H]=--1.3
and 2 km s$^{\rm -1}$, respectively.  Microturbulence was further refined by
removing trends in Fe I abundance as a function of line strength.  The final
model metallicity was set as the average [Fe/H] value derived from the Fe I
lines.  Our adopted model atmosphere parameters, along with available 
photometry, star identifications, and radial velocities, are provided in 
Table 1.

\subsection{Derivation of Iron, Sodium, and Oxygen Abundances}

Iron abundances were determined by measuring equivalent widths (EWs) for up
to 25 Fe I lines, using the line list from Johnson et al. (2014, submitted).  
The EWs were measured using the fitting code developed for Johnson et al. 
(2008), which fits either a single Gaussian or can deblend up to five nearby 
Gaussian profiles.  The abundances were determined using the 2010 version of 
the LTE line analysis code MOOG (Sneden 1973).  A summary of the average [Fe/H] 
abundances is provided in Table 1.

For the measurement of sodium and oxygen abundances, we utilized the spectrum
synthesis driver in MOOG and created synthetic spectra for each star in the
wavelength regions 6150--6165 \AA\ for Na and 6295--6305 \AA\ for oxygen.  
The abundances of the other elements for each star were held constant, except
the nitrogen abundance was used as a proxy to fit CN lines.  We adopted the 
O and Na line lists from Johnson et al. (2014, submitted).  The final 
abundances were set by fitting the smoothed synthetic spectrum to the observed 
spectrum by eye.  Typical fitting uncertainties are $\sim$0.10 dex.  The [O/Fe]
and [Na/Fe] ratios are provided in Table 1.

\subsection{Abundance Uncertainties}

In order to estimate the sensitivity of our derived abundances ratios due to
uncertainties in the adopted model atmosphere parameters, we rederived the 
[Fe/H], [O/Fe], and [Na/Fe] ratios by varying one parameter at a time and 
holding the others fixed.  We estimated conservative uncertainties of 
T$_{\rm eff}$$\pm$100 K, log(g)$\pm$0.3 cgs, [M/H]$\pm$0.3 dex, and 
vt$\pm$0.15 km s$^{\rm -1}$.  The average total uncertainty for [Fe/H], [O/Fe],
and [Na/Fe] is calculated to be $\pm$0.15, $\pm$0.18, and $\pm$0.16 dex, 
respectively.  Since our iron and sodium abundances are based solely on the 
measurement of neutral lines, the derived [Fe/H] and [Na/Fe] values were
most sensitive to variations in effective temperature.  In contrast, the 
[O/Fe] ratios, especially in the absence of reliable Fe II measurements,
were most sensitive to uncertainties in the surface gravity.  Individual
errors, taking into account the line--to--line abundance dispersion and 
all model parameter uncertainties, are listed in Table 1.

\section{RESULTS AND DISCUSSION}

\subsection{Basic Composition Results}

Previous large sample ($\ga$10 stars), high resolution spectroscopic analyses 
of NGC 288 have found the cluster to be moderately metal--poor, with estimates 
ranging from $\langle$[Fe/H]$\rangle$=--1.22 ($\sigma$=0.04; Carretta et al. 
2009a) to $\langle$[Fe/H]$\rangle$=--1.39 ($\sigma$=0.04; Shetrone \& Keane 
2000).  We find in agreement with past work that NGC 288 has 
$\langle$[Fe/H]$\rangle$=--1.19 ($\sigma$=0.12).  Although the star--to--star
dispersion in [Fe/H] is larger than those reported in previous studies 
(e.g., Carretta et al. 2009a), this is likely due to the moderate S/N of our
data and does not signify the existence of an intrinsic metallicity spread
within the cluster.

\begin{figure}
\epsscale{1.00}
\plotone{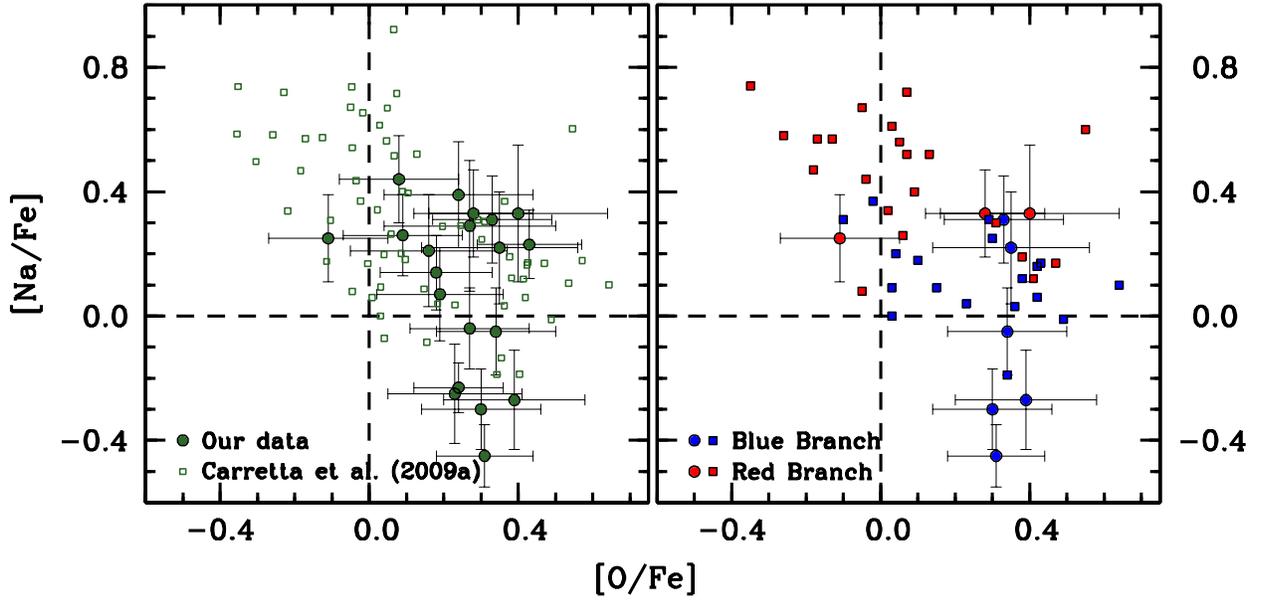}
\caption{\emph{left:} A plot of [Na/Fe] versus [O/Fe] for stars in the present
work (filled green circles) and Carretta et al. (2009a; open green boxes).
\emph{right:} A similar plot showing the O--Na anti--correlation in the 
present data (circles) and Carretta et al. (2009a; boxes), but the stars are 
color--coded depending on whether each lies on the blue or red RGB in 
Figure \ref{f1}.  Horizontal branch, AGB, and RGB--tip stars that cannot be
reliably assigned to either RGB in Figure \ref{f1} have been omitted in this
panel.  The red RGB stars have predominantly low [O/Fe] and high [Na/Fe] 
ratios.  In contrast, the blue RGB stars typically have comparatively high 
[O/Fe] and low [Na/Fe] ratios.}
\label{f2}
\end{figure}

The O--Na anti--correlation and CN/CH variations, which are common features of 
nearly all Galactic globular clusters (e.g., Gratton et al. 2004; 2012), has 
been previously reported in NGC 288 stars (e.g., Shetrone
\& Keane 2000; Kayser et al. 2008; Carretta et al. 2009a; 2009b; Smith \& 
Langland--Shula 2009).  The light element abundance variations 
in NGC 288 have also been linked to two populations having slightly different 
He abundances ($\Delta$Y=0.013; Piotto et al. 2013).  While we cannot directly 
test for He abundance variations with our current data set, we have measured
[O/Fe] (20/27 stars) and [Na/Fe] (26/27 stars) abundances.  As can be seen in
Figure \ref{f2}, we confirm the existence of the O--Na anti--correlation in
NGC 288.  We also find $\langle$[O/Fe]$\rangle$=$+$0.25 ($\sigma$=0.13) and 
$\langle$[Na/Fe]$\rangle$=$+$0.15 ($\sigma$=0.26), which is in reasonable 
agreement with past work by Carretta et al. (2009a;
$\langle$[O/Fe]$\rangle$=$+$0.11; $\sigma$=0.26; 
$\langle$[Na/Fe]$\rangle$=$+$0.27; $\sigma$=0.26) and Shetrone \& Keane (2000;
$\langle$[O/Fe]$\rangle$=$+$0.22; $\sigma$=0.14;  
$\langle$[Na/Fe]$\rangle$=$+$0.20; $\sigma$=0.25).  

Previous work on the CN and CH variations of sub--giant branch stars
(e.g., Kayser et al. 2008; Smith \& Langland--Shula 2009), as well as the 
O, Na, and Al abundances in RGB stars (e.g., Shetrone \& Keane 2008; Carretta 
et al. 2009a; 2009b), found some indications that the light element chemistry 
may be distributed in a bimodal manner.  While the [O/Fe] and [Na/Fe] data 
presented here do not exhibit a strongly bimodal distribution, possibly 
because of the lower S/N of our spectra and modest sample size, such a 
distribution may provide an important clue in understanding the split RGB 
reported in Roh et al. (2011).

\subsection{A New CTIO Ca Filter Tracing and Linking the Split RGB to Chemical
Composition}

Although Roh et al. (2011) initially ruled out contamination of the CTIO Ca 
filter from the CN band at $\sim$3885 \AA, based on previous filter tracing
data, the authors noted that the colors of the split RGB in NGC 288 were 
correlated with the cyanogen index measurements of Kayser et al. (2008).  In
particular, they found that the ``CN--normal" (i.e., halo--like composition)
stars were located on the blue RGB but the ``CN--strong" stars were located on
the red RGB (see Figure 4 of Roh et al. 2011).  If real, the chemical 
distinction between the two branches should also be detectable on the upper RGB
as differences in [O/Fe] and [Na/Fe] abundances because CN--normal and 
CN--strong stars are expected to be O--rich/Na--poor and O--poor/Na--rich, 
respectively.

In Figure \ref{f2} we investigated the correlation between RGB color and 
light element abundances, using both our data and those from Carretta et al.
(2009a), and found the same correlation between color and composition.  While
the dispersion in both [O/Fe] and [Na/Fe] is roughly a factor of two in 
magnitude, the blue RGB is clearly dominated by stars with higher [O/Fe] and 
lower [Na/Fe] compared to the red RGB.  On average, the blue RGB stars from
the combined data set have $\langle$[O/Fe]$\rangle$=$+$0.27 ($\sigma$=0.19)
and $\langle$[Na/Fe]$\rangle$=$+$0.11 ($\sigma$=0.16) while the red RGB stars
have $\langle$[O/Fe]$\rangle$=$+$0.08 ($\sigma$=0.24) and 
$\langle$[Na/Fe]$\rangle$=$+$0.43 ($\sigma$=0.19).  In contrast, the difference
in [Fe/H] between the two branches is mostly negligible, with the blue RGB
having $\langle$[Fe/H]$\rangle$=--1.19 ($\sigma$=0.08) and the red RGB having
$\langle$[Fe/H]$\rangle$=--1.24 ($\sigma$=0.06).  Therefore, we conclude that
the split RGB in NGC 288 is driven almost exclusively by differences in
light element composition (and possibly He) rather than a difference of 0.16 
dex in [M/H] and 1.5 Gyr in age, as suggested by Roh et al. (2011).

\begin{figure}
\epsscale{1.00}
\plotone{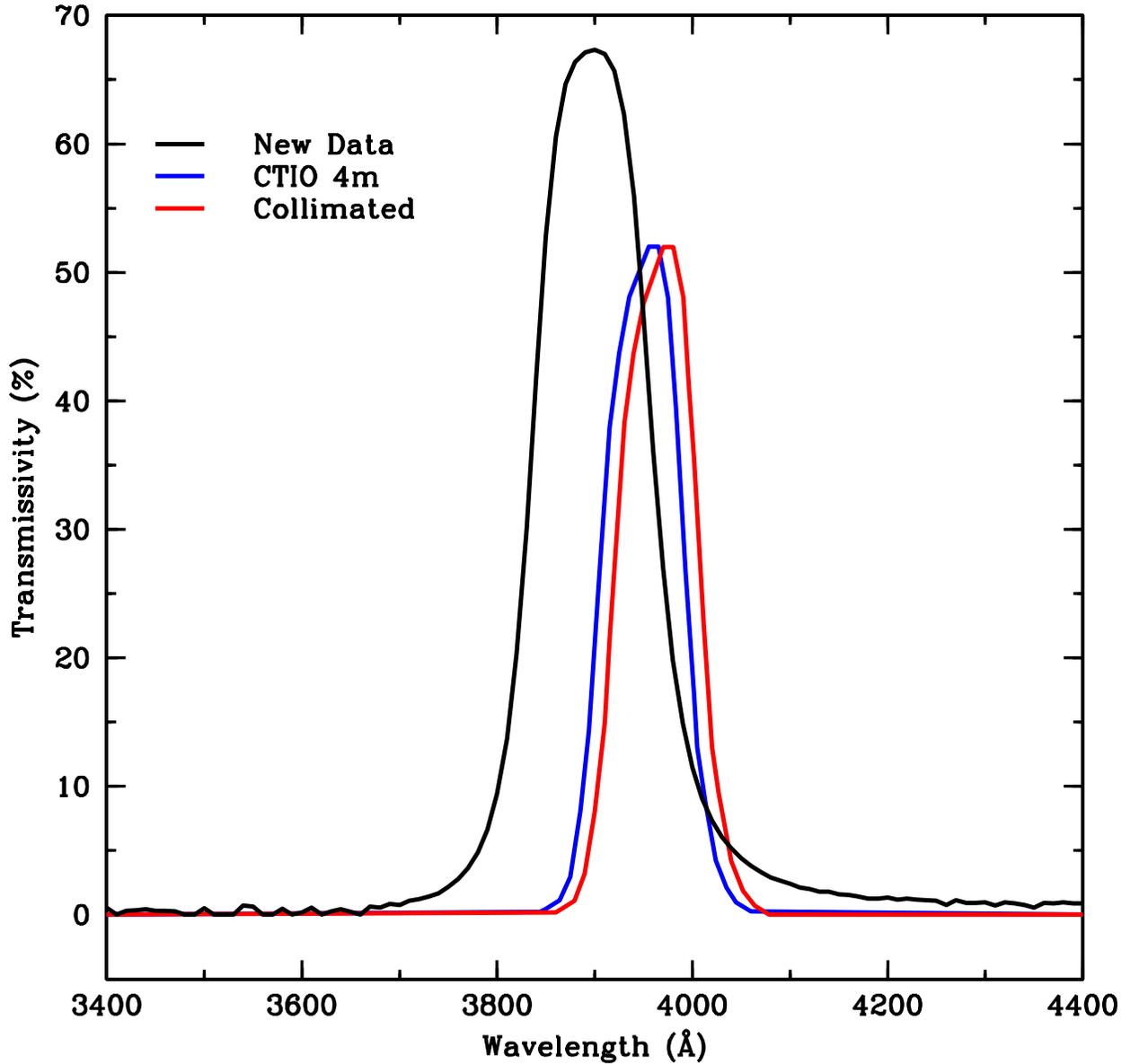}
\caption{The old filter response curve for the CTIO Ca filter, including for 
both a collimated beam (red line) and mounted at prime focus (blue line), are
compared with the newly measured filter response function (black line).  Note 
in particular that the new lab measurement finds the peak transmission to occur
near 3900 \AA, which is $>$50 \AA\ bluer than the original curve.  When 
compared with Figure 4 in Roh et al. (2011), it is clear that the CTIO Ca 
filter is contaminated by the CN--band blueward of $\sim$3885 \AA.}
\label{f3}
\end{figure}

The remaining question is \emph{why} the data from Roh et al. (2011) show a
double RGB when the CTIO Ca filter is supposed to avoid major contamination 
from nearby molecular CN bands.  To investigate this further, we requested a 
new transmission curve of the CTIO Ca narrow--band filter.  The filter 
response was measured using a Cary 500 spectrophotometer at Gemini 
Observatory with a collimated beam at incident angle.  The new filter response
curve (A. Kunder \& D. H{\"o}lck 2012, private communication) is shown in 
Figure \ref{f3} (see also Table 2), along with the original data provided to 
Roh et al. (2011).  It is clear from Figure \ref{f3} that the CTIO Ca filter 
actually peaks in transmissivity near 3900 \AA, and thus can be significantly 
contaminated by the CN band blueward of $\sim$3885 \AA\ in cool stars.  As
mentioned in Lee et al. (2009) and Roh et al. (2011), the CN contamination is 
likely to be increased when the filter is placed at prime focus and the peak 
transmission shifts blueward.  This is the root cause of the split RGB found 
by Roh et al. (2011).  A new filter, specifically designed to trace only the 
Ca HK region without significant CN contamination and used with the du Pont 
2.5m at Las Campanas Observatory, confirms our result as the double RGB in 
NGC 288 disappears in updated y vs hk color--magnitude diagrams (Lee et al. 
2014, in prep.; see also Lee et al 2013).  Interestingly, although the CTIO Ca 
filter may be too contaminated to use with the Ca HK lines, it may be an 
efficient filter for separating CN--normal and CN--strong stars in other 
stellar populations.

In addition to NGC 288, the CN contamination also likely explains the RGB hk 
index spreads in the other clusters (NGC 2808, M4, M5, NGC 6752, and NGC 6397)
observed by Lee et al. (2009) for which significant [Ca/H] and/or [Fe/H] 
variations are not confirmed by high--resolution spectroscopy (e.g., Carretta
et al. 2010).  However, the RGB hk index spreads are likely real in the 
remaining clusters $\omega$ Cen, M22, and NGC 1851 (e.g., Lee et al. 2009; 
Joo \& Lee 2013), which have spectroscopically confirmed star--to--star [Ca/H] 
and/or [Fe/H] variations.  Finally, we note that while Yong et al. (2013) find
a statistically significant correlation between Na and Ca abundances in 
NGC 6752, the amplitudes of the star--to--star abundance variations are at the 
few percent level.  This effect, if also present in a similarly monometallic
cluster such as NGC 288, is likely not strong enough to contribute 
significantly to the broadly separated color sequences seen in the Roh et al.
(2011) y vs hk color--magnitude diagrams.

\section{SUMMARY}

We derived [Fe/H], [O/Fe], and [Na/Fe] abundances for 27 RGB stars in the 
Galactic globular cluster NGC 288 using high resolution spectra obtained 
with the Hydra multifiber spectrograph on the Blanco 4m telescope at CTIO.
We find that the cluster has a negligible spread in metallicity with an 
average [Fe/H]=--1.19 ($\sigma$=0.12).  We also confirm the existence of the
O--Na anti--correlation in NGC 288 RGB stars.  The average oxygen and sodium
abundances from our sample are $\langle$[O/Fe]$\rangle$=$+$0.25 ($\sigma$=0.13)
and $\langle$[Na/Fe]$\rangle$=$+$0.15 ($\sigma$=0.26), respectively.  
Furthermore, we find that the two giant branches previously noted by Roh et 
al. (2011) are driven by differences in the abundances of elements from
carbon through sodium (and probably also magnesium and aluminum), rather than
a combination of age and metallicity spreads.  In general, the blue RGB stars 
have significantly higher [O/Fe] and lower [Na/Fe] than the red RGB stars.
Further investigation of the double RGB lead us to discover that this feature 
was due primarily to contamination by a CN--band blueward of 3885 \AA\ in the
original data set.  A new filter response curve, obtained for this study, 
revealed the peak transmission to be $>$50 \AA\ bluer than originally thought,
and is thus the real cause of the double RGB found in Roh et al. (2011).

\acknowledgements

We thank the referee for a careful review and thoughtful comments that lead
to an improvement of the manuscript.  This publication makes use of data 
products from the Two Micron All Sky Survey, which is a joint project of the 
University of Massachusetts and the Infrared Processing and Analysis 
Center/California Institute of Technology, funded by 
the National Aeronautics and Space Administration and the National Science 
Foundation.  We thank the Cerro Tololo Inter--American Observatory (CTIO) staff
Daniel H{\"o}lck and Andrea Kunder for providing the new 
response curve for the CTIO Ca filter.  CIJ gratefully acknowledges support 
from the Clay Fellowship, administered by the Smithsonian Astrophysical 
Observatory.  This material is based upon work supported by the National 
Science Foundation under award No. AST--1003201 to CIJ.  RMR acknowledges 
support from NSF grants AST--0709479 and AST--12112099.  Y.--W. L. acknowledges
support from NRF of Korea to CGER.

\clearpage

\clearpage

\clearpage

\clearpage
\setlength{\voffset}{0.90in}
\tablenum{1}
\tablecolumns{16}
\tablewidth{0pt}

\begin{deluxetable}{cccccccccccccccc}
\tabletypesize{\scriptsize}
\rotate
\tablecaption{Star Identifications, Coordinates, Model Atmosphere
Parameters, and Radial Velocities}
\tablehead{
\colhead{ID}	&
\colhead{RA (J2000)}	&
\colhead{DEC (J2000)}      &
\colhead{J}      &
\colhead{H}      &
\colhead{K$_{\rm S}$}      &
\colhead{T$_{\rm eff}$}      &
\colhead{log(g)}      &
\colhead{[Fe/H]}      &
\colhead{vt}      &
\colhead{[O/Fe]}	&
\colhead{[Na/Fe]}        &
\colhead{[Fe/H]}      &
\colhead{[O/Fe]}        &
\colhead{[Na/Fe]}        &
\colhead{RV$_{\rm helio.}$} \\
\colhead{2MASS}	&
\colhead{(degrees)}      &
\colhead{(degrees)}      &
\colhead{(mag.)}      &
\colhead{(mag.)}      &
\colhead{(mag.)}      &
\colhead{(K)}      &
\colhead{(cgs)}      &
\colhead{}      &
\colhead{(km s$^{\rm -1}$)}      &
\colhead{}      &
\colhead{}      &
\colhead{Error}      &
\colhead{Error}      &
\colhead{Error}      &
\colhead{(km s$^{\rm -1}$)}      
}
\startdata
00521548$-$2641039	&	13.064524	&	$-$26.684423	&	13.366	&	12.815	&	12.816	&	5015	&	2.15	&	$-$0.94	&	1.60	&	\nodata	&	$+$0.54	&	0.14	&	\nodata	&	0.15	&	$-$42.1	\\
00522151$-$2632418	&	13.089655	&	$-$26.544960	&	11.221	&	10.578	&	10.402	&	4225	&	1.05	&	$-$1.13	&	1.70	&	$+$0.23	&	$-$0.25	&	0.15	&	0.18	&	0.16	&	$-$43.8	\\
00523129$-$2639008	&	13.130410	&	$-$26.650232	&	13.098	&	12.693	&	12.601	&	5215	&	2.10	&	$-$1.23	&	2.00	&	$+$0.43	&	$+$0.23	&	0.10	&	0.14	&	0.11	&	$-$39.7	\\
00523149$-$2636016	&	13.131247	&	$-$26.600452	&	12.881	&	12.320	&	12.234	&	4690	&	1.85	&	$-$1.19	&	1.40	&	$+$0.24	&	$+$0.39	&	0.17	&	0.20	&	0.17	&	$-$38.8	\\
00523688$-$2630586	&	13.153706	&	$-$26.516281	&	12.572	&	11.985	&	11.918	&	4665	&	1.75	&	$-$1.01	&	1.60	&	$+$0.16	&	$+$0.21	&	0.18	&	0.21	&	0.18	&	$-$41.9	\\
00523795$-$2636064	&	13.158162	&	$-$26.601801	&	13.062	&	12.560	&	12.424	&	4715	&	1.95	&	$-$1.04	&	1.30	&	$+$0.39	&	$-$0.27	&	0.16	&	0.19	&	0.16	&	$-$41.1	\\
00523890$-$2635317	&	13.162090	&	$-$26.592154	&	11.609	&	10.923	&	10.802	&	4255	&	1.25	&	$-$1.23	&	1.50	&	$+$0.30	&	$-$0.30	&	0.12	&	0.16	&	0.13	&	$-$46.1	\\
00523998$-$2633498	&	13.166596	&	$-$26.563854	&	11.842	&	11.195	&	11.086	&	4380	&	1.35	&	$-$1.09	&	1.50	&	$+$0.24	&	$-$0.23	&	0.07	&	0.12	&	0.08	&	$-$42.2	\\
00524033$-$2636328	&	13.168067	&	$-$26.609137	&	13.621	&	13.112	&	13.074	&	5025	&	2.25	&	$-$1.13	&	2.00	&	$+$0.40	&	$+$0.33	&	0.22	&	0.24	&	0.22	&	$-$45.4	\\
00524057$-$2628033	&	13.169044	&	$-$26.467602	&	13.438	&	12.901	&	12.839	&	4840	&	2.10	&	$-$1.29	&	1.90	&	$+$0.09	&	$+$0.26	&	0.12	&	0.16	&	0.13	&	$-$41.7	\\
00524095$-$2634390	&	13.170638	&	$-$26.577507	&	12.771	&	12.248	&	12.170	&	4835	&	1.85	&	$-$1.01	&	1.45	&	$+$0.34	&	$-$0.05	&	0.13	&	0.16	&	0.14	&	$-$41.0	\\
00524379$-$2638018	&	13.182459	&	$-$26.633837	&	11.715	&	11.151	&	11.045	&	4620	&	1.40	&	$-$1.32	&	1.80	&	$+$0.19	&	$+$0.07	&	0.14	&	0.17	&	0.15	&	$-$41.8	\\
00524526$-$2633297	&	13.188614	&	$-$26.558271	&	12.805	&	12.381	&	12.280	&	5105	&	1.95	&	$-$1.36	&	2.00	&	\nodata	&	$+$0.41	&	0.22	&	\nodata	&	0.22	&	$-$42.0	\\
00524625$-$2635494	&	13.192748	&	$-$26.597057	&	11.410	&	10.729	&	10.622	&	4300	&	1.15	&	$-$1.33	&	1.75	&	$+$0.28	&	$+$0.33	&	0.13	&	0.16	&	0.14	&	$-$39.3	\\
00524655$-$2631523	&	13.193994	&	$-$26.531214	&	13.520	&	12.940	&	12.904	&	4785	&	2.15	&	$-$1.24	&	1.80	&	\nodata	&	$+$0.09	&	0.29	&	\nodata	&	0.29	&	$-$42.0	\\
00524910$-$2640072	&	13.204609	&	$-$26.668690	&	13.237	&	12.661	&	12.596	&	4705	&	2.00	&	$-$1.21	&	1.40	&	$+$0.33	&	$+$0.31	&	0.13	&	0.16	&	0.14	&	$-$44.1	\\
00524957$-$2642106	&	13.206557	&	$-$26.702971	&	11.788	&	11.176	&	11.050	&	4430	&	1.35	&	$-$1.17	&	1.65	&	$+$0.27	&	$-$0.04	&	0.12	&	0.16	&	0.13	&	$-$48.7	\\
00524958$-$2637006	&	13.206591	&	$-$26.616846	&	13.741	&	13.235	&	13.143	&	4845	&	2.25	&	$-$1.26	&	1.55	&	\nodata	&	$+$0.21	&	0.11	&	\nodata	&	0.12	&	$-$42.7	\\
00524977$-$2635118	&	13.207381	&	$-$26.586620	&	12.546	&	11.970	&	11.908	&	4715	&	1.75	&	$-$1.29	&	1.75	&	$+$0.08	&	$+$0.44	&	0.13	&	0.16	&	0.14	&	$-$42.2	\\
00524981$-$2636559	&	13.207542	&	$-$26.615547	&	12.293	&	11.675	&	11.587	&	4515	&	1.60	&	$-$1.41	&	1.75	&	\nodata	&	$+$0.41	&	0.20	&	\nodata	&	0.20	&	$-$49.4	\\
00525240$-$2632252	&	13.218359	&	$-$26.540346	&	13.758	&	13.285	&	13.195	&	4965	&	2.30	&	$-$1.20	&	1.95	&	\nodata	&	$+$0.30	&	0.15	&	\nodata	&	0.16	&	$-$42.0	\\
00525279$-$2634388	&	13.219980	&	$-$26.577454	&	11.510	&	10.798	&	10.674	&	4185	&	1.20	&	$-$1.11	&	1.50	&	$+$0.31	&	$-$0.45	&	0.09	&	0.13	&	0.10	&	$-$52.2	\\
00525287$-$2635201	&	13.220298	&	$-$26.588942	&	10.905	&	10.197	&	10.075	&	4200	&	0.95	&	$-$1.32	&	2.00	&	$+$0.27	&	$+$0.29	&	0.21	&	0.23	&	0.21	&	$-$56.2	\\
00525460$-$2637084	&	13.227506	&	$-$26.619019	&	12.275	&	11.780	&	11.648	&	4750	&	1.65	&	$-$1.30	&	1.90	&	$+$0.18	&	$+$0.14	&	0.11	&	0.15	&	0.12	&	$-$45.4	\\
00525842$-$2637498	&	13.243453	&	$-$26.630505	&	13.762	&	13.185	&	13.176	&	4885	&	2.25	&	$-$1.05	&	1.35	&	$+$0.35	&	$+$0.22	&	0.18	&	0.21	&	0.18	&	$-$38.7	\\
00531301$-$2637047	&	13.304216	&	$-$26.617979	&	12.888	&	12.305	&	12.226	&	4645	&	1.85	&	$-$1.20	&	1.80	&	$-$0.11	&	$+$0.25	&	0.13	&	0.16	&	0.14	&	$-$43.0	\\
00531493$-$2633524	&	13.312218	&	$-$26.564575	&	13.363	&	12.820	&	12.768	&	4855	&	2.10	&	$-$1.20	&	1.65	&	\nodata	&	\nodata	&	0.21	&	\nodata	&	\nodata	&	$-$41.5	\\
\enddata

\end{deluxetable}

\clearpage
\tablenum{2}
\tablecolumns{2}
\tablewidth{0pt}

\begin{deluxetable}{cc}
\tablecaption{CTIO Ca Filter Response}
\tablehead{
\colhead{Wavelength}    &
\colhead{Transmission}	\\
\colhead{(\AA)}	&
\colhead{(\%)}
}

\startdata
3400	&	0.55	\\
3410	&	0.00	\\
3420	&	0.30	\\
3430	&	0.36	\\
3440	&	0.42	\\
3450	&	0.30	\\
3460	&	0.30	\\
3470	&	0.25	\\
3480	&	0.00	\\
3490	&	0.00	\\
3500	&	0.51	\\
3510	&	0.00	\\
3520	&	0.00	\\
3530	&	0.00	\\
3540	&	0.72	\\
3550	&	0.65	\\
3560	&	0.00	\\
3570	&	0.00	\\
3580	&	0.46	\\
3590	&	0.00	\\
3600	&	0.20	\\
3610	&	0.55	\\
3620	&	0.00	\\
3630	&	0.25	\\
3640	&	0.42	\\
3650	&	0.22	\\
3660	&	0.00	\\
3670	&	0.62	\\
3680	&	0.57	\\
3690	&	0.83	\\
3700	&	0.76	\\
3710	&	1.10	\\
3720	&	1.22	\\
3730	&	1.42	\\
3740	&	1.69	\\
3750	&	2.16	\\
3760	&	2.80	\\
3770	&	3.60	\\
3780	&	4.84	\\
3790	&	6.63	\\
3800	&	9.38	\\
3810	&	13.68	\\
3820	&	20.32	\\
3830	&	30.04	\\
3840	&	41.90	\\
3850	&	52.84	\\
3860	&	60.57	\\
3870	&	64.63	\\
3880	&	66.39	\\
3890	&	67.13	\\
3900	&	67.34	\\
3910	&	67.00	\\
3920	&	65.67	\\
3930	&	62.35	\\
3940	&	55.92	\\
3950	&	46.48	\\
3960	&	36.17	\\
3970	&	26.90	\\
3980	&	19.86	\\
3990	&	14.90	\\
4000	&	11.44	\\
4010	&	9.07	\\
4020	&	7.35	\\
4030	&	6.10	\\
4040	&	5.17	\\
4050	&	4.39	\\
4060	&	3.81	\\
4070	&	3.31	\\
4080	&	2.91	\\
4090	&	2.68	\\
4100	&	2.42	\\
4110	&	2.14	\\
4120	&	2.01	\\
4130	&	1.80	\\
4140	&	1.79	\\
4150	&	1.61	\\
4160	&	1.53	\\
4170	&	1.42	\\
4180	&	1.28	\\
4190	&	1.27	\\
4200	&	1.33	\\
4210	&	1.19	\\
4220	&	1.26	\\
4230	&	1.17	\\
4240	&	1.15	\\
4250	&	1.10	\\
4260	&	0.78	\\
4270	&	1.13	\\
4280	&	0.94	\\
4290	&	0.95	\\
4300	&	0.97	\\
4310	&	0.73	\\
4320	&	0.97	\\
4330	&	0.88	\\
4340	&	0.74	\\
4350	&	0.54	\\
4360	&	0.94	\\
4370	&	0.88	\\
4380	&	0.96	\\
4390	&	0.87	\\
4400	&	0.91	\\
\enddata

\end{deluxetable}

\end{document}